\documentstyle[twoside,fleqn,espcrc2]{article}

\newbox\SlashedBox  
\def\fs#1{\setbox\SlashedBox=\hbox{#1} 
\hbox to 0pt{\hbox to 1\wd\SlashedBox{\hfil/\hfil}\hss}{#1}} 
\def\hboxtosizeof#1#2{\setbox\SlashedBox=\hbox{#1} 
\hbox to 1\wd\SlashedBox{#2}} 

\def\ms#1{\setbox\SlashedBox=\hbox{$#1$}
\hbox to 0pt{\hbox to 1\wd\SlashedBox{\hfil/\hfil}\hss}#1}

\newcommand{\tr}{{\rm tr}}

\newcommand{\ie}{{\em i.e.~}}

\newcommand{\ap}{{\alpha^{\prime}}}

\newcommand{\be}{\begin{equation}}
\newcommand{\ee}{\end{equation}}
\newcommand{\ba}{\begin{eqnarray}}
\newcommand{\ea}{\end{eqnarray}}
\newcommand{\bea}{\begin{eqnarray}}
\newcommand{\eea}{\end{eqnarray}}

\newcommand{\AmS}{{\protect\the\textfont2
  A\kern-.1667em\lower.5ex\hbox{M}\kern-.125emS}}

\begin{document}
    
\title{(Non-)perturbative tests of the AdS/CFT correspondence}

\author{Massimo BIANCHI\thanks{On leave of absence from Dipartimento di Fisica, 
Universit\`a di Roma ``Tor Vergata'', Via della Ricerca 
Scientifica, Roma 00133, ITALY.} \\
Department of Applied Mathematics 
and Theoretical Physics,\\ 
University of Cambridge, Centre for Mathematical Sciences, \\ 
Wilberforce Road, CB3 0WA Cambridge, England}

\begin{abstract}
{I summarize perturbative and non-perturbative field theory tests of the 
holographic correspondence between 
type IIB superstring on $AdS_{5}\times S^{5}$ and ${\cal N}=4$ SYM 
theory. The holographic duality between D-instantons and YM instantons is 
briefly described. Non renormalization of two- and three-point functions of 
CPO's and their extremal and next-to-extremal correlators are then reviewed.
Finally, partial non-renormalization of four-point functions of 
lowest CPO's is analyzed in view of the interpretation of short 
distance logarithmic behaviours in terms
of anomalous dimensions of unprotected operators.}
\vspace{1pc}
\end{abstract}
\maketitle

The AdS/CFT correspondence \cite{JM} is
an unprecedented tool in the study of the interplay between gauge 
theory and
gravity. In the simplest  
case it
relates type IIB superstring on $AdS_5\times S^5$ to ${\cal N}=4$ 
supersymmetric
Yang-Mills theory (SYM) with gauge group $SU(N)$. The 
correspondence
is ``holographic'' in that the gauge theory lives on the boundary of AdS. 
The conjecture is motivated by the study of the low-energy
dynamics of D-branes \cite{MAGOO}. Their open 
strings excitations include massless vector supermultiplets and 
a stack of $N$ coincident D3-branes is governed by $U(N)$ 
${\cal N}=4$ SYM in $D=4$.  The remarkable fact 
about this gauge theory is its exact
superconformal invariance. This reflects into the 
the dilaton being constant and the metric being nowhere singular 
for the D3-brane solution!
The D3-brane can thus be viewed as a smooth 
soliton interpolating between maximally supersymmetric 
flat Minkowski spacetime at infinity and 
maximally supersymmetric $AdS_5\times S^5$ 
near the horizon. 
The AdS scale $L$ is related to the RR 5-form flux $N$ by
$L^4 = 4\pi g_s N \ap^2$. Due to the red-shift, the geometry 
near the horizon captures the low-energy limit
where bulk supergravity effectively decouples from the 
boundary dynamics. 
Open-closed string duality suggests a 
perfect equivalence between the two descriptions should 
take place when 
\be 
g_s = \langle e^\Phi \rangle = {g^2_{_{YM}} \over 4\pi}
\qquad 
\vartheta_s = \langle \chi \rangle = {\theta \over 2\pi} 
\label{dictio} 
\ee
and $N$ is identified with the number of colours.
The type IIB ``partition function''
plays the role of a generating functional 
\be 
Z_{IIB}[ \Phi[J] ] = Z_{SYM} [J] \; .
\ee 
The boundary data $J(x)$ for type IIB bulk fields $\Phi_{M}(x,\rho)$
are viewed as sources for gauge-invariant SYM local composite 
operators ${\cal{O}}_{\Delta}(x)$. Since string theory on  
AdS spaces with RR background is still poorly understood even at the
classical level, it is necessary to perform 
a double expansion in powers of the string
coupling $g_{s}$ and inverse tension $\ap$ or, equivalently,
in powers of $1/N$ and $1/\kappa$, where $\kappa = g_{YM}^{2} N$ is 
the reduced `t Hooft coupling. Taking $N$ to infinity one is effectively 
neglecting string loops and restricting to the tree level (sphere)  dual 
to the ``planar series''.
This is still insufficient for computational 
purposes. In order to suppress 
higher derivative corrections one has to restrict to large $\kappa$ 
\ie to the strong coupling regime in the dual gauge theory.
In principle, one can systematically take into account
worldsheet corrections, string-loops and non-perturbative
D-instanton corrections \cite{GG}. 

A string description of confinement has long been sought for.
Just before Maldacena's proposal, Alexander Polyakov observed 
that one of
the drawbacks of previous attempts, \ie the lack of {\it 
zig-zag} symmetry of the non-critical string,  
could be cured by assuming the flow to a 
fixed point at vanishing Liouville field $\rho = 0$ \cite{AMP}.
Maldacena's proposal then looks like what the 
Doctor orders
in that it puts forward the existence of a fifth coordinate $\rho$ 
(transverse to
the boundary) that could be identified with the Liouville mode or, 
possibly equivalently, with a renormalization scale. 
What sounds surprising is that ${\cal N} =4$ SYM 
theory is not confining and has no mass-gap in the superconformal phase. 

The superisometry group of $AdS_5\times  S_5$, $SU(2,2|4)$, acts 
by superconformal transformations on the boundary CFT. 
Unitary irreducible representations of $PSU(2,2|4)$ are 
labelled by the
quantum numbers $\{\Delta, j_L, j_R; [k,l,m]\}$. 
$\Delta$ is the scaling dimension, $(j_L, j_R)$ denote the 
$SU(2)_{L}\times SU(2)_{R}$ spins, and $[k,l,m]$ are the Dynkin labels of
an irrep ${\bf r}$ of the $SU(4)$ R-symmetry group. 
The fundamental SYM fields $\{\varphi^{i}, \lambda^{A}, F_{\mu\nu}\}$
describing the lowest lying open-string excitations belong to the 
singleton representation and live on the boundary. Gauge invariant 
composite operators ${\cal O}_{\Delta}$ dual to
type IIB bulk fields $\Phi_{M}$ form doubleton or long multiplets. In 
particular, the ``ultra-short'' 
${\cal N}=4$ supercurrent multiplet is dual to the
``massless'' AdS supergravity multiplet. Operators dual to
higher KK excitations 
assemble into short multiplets 
with spin $j = j_L + j_R \leq 2$ and Dynkin 
labels $k,m\leq 2$. 
Their lowest components are chiral primary operators (CPO's) 
\begin{displaymath}
    {\cal Q}^{i_{1}i_{2}\ldots i_{\ell}}_{[0,\ell,0]}=\sum_{\sigma} 
    Tr[\varphi^{\sigma(i_{1})} \varphi^{\sigma(i_{2})}\ldots 
    \varphi^{\sigma(i_{\ell})} - ...]
\nonumber
\end{displaymath}
of dimension $\Delta = \ell$ belonging to the $\ell$-fold  
traceless symmetric product 
of the ${\bf 6}$ of $SU(4)\approx SO(6)$.
Other shortenings are possible for multi-trace operators. For scalar 
primaries, for instance, shortening occurs when ${\bf r} = 
[k,\ell,k]$ and
$\Delta=\ell +2k$ or ${\bf r} = [k+2n,\ell,k]$ and $\Delta=\ell + 2k 
+ 3n$ \cite{AFZ}. The spin ranges over 6 or 7 units respectively.
Operators dual to string excitations with AdS masses of order
$1/\sqrt{\ap}$
belong to long multiplets. Their spin ranges over 8 units
and their dimension is expected to grow like $\Delta 
\approx
\kappa^{1/4}$ in the strong coupling limit. One such example is the 
${\cal N}=4$ Konishi multiplet whose lowest component is the scalar $SU(4)$  
singlet
\be 
{\cal K}_{\bf 1} = Tr (\varphi^i \varphi_{i}) 
\ee
with naive dimension $\Delta=2$. 

In addition to perturbative symmetries, both type IIB superstring and 
${\cal N}=4$ SYM are expected to be invariant under S-duality.
Charge quantization of solitonic states breaks the 
non-compact $SL(2,R)$ symmetry of ``classical'' type IIB supergravity to 
$SL(2,Z)$. The breaking is manifest in the 
D-istanton corrections to the type IIB effective action \cite{GG}, that
are dual to SYM instanton corrections \cite{BG,EWB,BGKR}.
The Type IIB D3-brane soliton is self-dual,
while strings, 5-branes and 7-branes form multiplets of the
discrete non-compact symmetry.
Similarly, stable dyonic states of ${\cal N}=4$ SYM transform 
into one another under $SL(2,Z)$. For simply laced gauge groups  
the theory is expected to be self-dual since the spectra of 
electric and magnetic charges coincide. For non-simply laced groups
S-duality maps one into the other. Theories with orthogonal and 
symplectic groups are dual to the near horizon geometry of D3-brane 
configurations in the presence of unorientifold planes with quantized 
possibly vanishing two form background \cite{BPS,EWB}.

Finally, three 
$U(1)$'s play an interesting subtle role in the correspondence.
The first, $U(1)_{Z}$,  
is a central extension of $PSU(2,2|4)$.  Fundamental fields as well as their 
composites are neutral with respect to it so that one usually neglects 
it. It is conceivable that solitonic states could carry non-vanishing 
$U(1)_{Z}$ charge and form novel $SU(2,2|4)$ multiplets \cite{GZ}.
The second, $U(1)_{C}$, is the abelian factor in $U(N)$. 
{From} the D3-brane 
perspective it corresponds to the center of mass degrees of freedom.
Its low-energy 
dynamics on the boundary cannot be reproduced by 
the bulk supergravity action. In the supergravity limit 
one could simply say there is an additional singleton multiplet 
not captured by the correspondence if not for its 
contribution to ``boundary anomalies'' \cite{BC}. 
When higher-derivative Born-Infeld corrections 
take over the 
theory is better described in terms of open strings.
The third, $U(1)_{B}$, is a ``bonus'' 
symmetry of a 
restricted class of correlation functions and their dual amplitudes
\cite{KI}. 
In SYM it corresponds to a chiral 
rotation accompanied by a continuous electric-magnetic duality 
transformation. 
Its type IIB counterpart is the $U(1)_{B}$ anomalous chiral symmetry. When 
supergravity loops and higher derivative string corrections are 
negligible the ``bonus'' symmetry becomes a true symmetry. 
Independently of the coupling $\kappa$ and $N$, 
all two-point correlation functions, three-point functions with at 
most one insertion of unprotected operators and four-point functions of 
single-trace protected operators 
seem to respect this symmetry \cite{KI}. 

In a superconformal field 
theory,
two-point functions of normalized primary operators ${\cal O}_\Delta$ 
are
completely specified by their dimensions 
\be 
\langle {\cal O}^{\dagger}_\Delta (x) {\cal
O}_\Delta (y)\rangle = {1\over (x-y)^{2\Delta}}
\label{twopoint}
\ee
The first step in computing two-point
functions of (scalar) gauge-invariant composite operators using the 
correspondence is to
solve the linearized field equation \cite{GKP,EWH,FMMR}
\be 
-\nabla^2 \Phi_{M} + M^2 \Phi_{M} = 0 
\ee 
with near-boundary behaviour\footnote{The other possible near-boundary behaviour 
$\Phi(\rho, x) \rightarrow\rho^{\Delta} V(x)$ corresponds to turning 
on a VEV $\langle {\cal O}\rangle = V$ for the operator dual to $\Phi$.} 
\be 
\Phi_{M}(\rho, x) \rightarrow \rho^{4-\Delta} J(x)   
\ee 
as $\rho \rightarrow 0$. The
solution may be expressed in terms of the bulk-to-boundary 
propagator   
\be 
K_{\Delta} (\rho, x; x') = 
{a_{\Delta} \rho^{\Delta}  \over (\rho^{2} + (x-x')^{2})^{\Delta} } 
\ee 
It is not at all a coincidence that $K_{\Delta}$ 
resembles a YM instanton form factor. 
Plugging  $K_{\Delta}$ into the scalar field equation one finds the following 
mass-to-dimension relation 
\be  
(ML)^2 = \Delta (\Delta - 4) 
\ee 
and its inverse 
\be 
\Delta = 2 \pm \sqrt{ 4 + (ML)^2} 
\ee 
$\Delta$ is real, as expected in a unitary theory, once the Breitenlohner 
Freedman bound $(ML)^{2}\geq - 4$ is enforced
\cite{BF}. Only the positive branch is relevant for ${\cal N}=4$ SYM. 
Carefully computing the quadratic on-shell type IIB action and differentiating 
wrt to the sources $J$ exactly 
reproduce the field theory result (\ref{twopoint}).
 
Three-point functions, although largely fixed by superconformal 
invariance, encode the dynamics of the theory since one can in principle
reconstruct all correlation functions by factorization. A particularly 
interesting class of three-point functions are those of CPO's
\be 
\langle Q_{\ell_1} (x_{1})  Q_{\ell_2} (x_{2}) Q_{\ell_3} (x_{3})
\rangle =  
{C({\ell_1}, {\ell_2}, {\ell_3}) \over 
\prod (x_{ij}^{2})^{{\ell_i}+{\ell_j}-{1\over 2}\Sigma} } 
\nonumber
\ee
where $x_{ij} = x_{i} - x_{j}$ and $\Sigma = {\ell_1} + {\ell_2} + {\ell_3}$.
The trilinear couplings
$C({\ell_1}, {\ell_2}, {\ell_3})$ 
can be easily computed at weak coupling for large $N$. 
In order to perform the dual AdS computation one has to go 
beyond the linearized approximation. Quadratic 
terms in the field equations, or equivalently cubic terms in the 
action are necessary. These are in general very complicated to extract
but the computation turns out to be 
feasible for CPO's. Quite remarkably,
one finds the same result as in free-field theory at large 
$N$ \cite{LMRS}. The exact matching suggests the validity of a 
non-renormalization theorem for any $\kappa$ and $N$. 
This has been tested at one-loop \cite{DFS} and to two-loops 
\cite{EHSSW2,BKRS2}. A judicious use of the ``bonus'' 
$U(1)_{B}$ symmetry \cite{KI} in 
the context of ${\cal N}$=2 harmonic superspace
gives a demonstration of the non-renormalization of two- and three-point
functions of CPO's~\cite{EHSSW3}. The extremal case, 
$\ell_{1} = \ell_{2} +\ell_{3}$, is subtler. We will return to 
this issue after discussing non-perturbative effects.

Using the AdS/CFT dictionary (\ref{dictio}), the charge-$k$  
type IIB D-instanton action  
coincides with the action of a charge-$k$ YM instanton. 
This strongly indicates a correspondence between 
these sources of non-perturbative effects \cite{BG,EWB}.
Moreover, it is known that the $k=1$ $SU(2)$ YM instanton moduli 
space coincides with Euclidean $AdS_{5}$ and the same is true for a 
type IIB D-instanton on $AdS_{5}$. The fifth radial 
coordinate transverse to the boundary plays the role of the YM instanton 
size $\rho$. The correspondence can be made more quantitative by 
noticing that the super-instanton measure contains an overall factor 
$g_{_{YM}}^8$ that arises from the combination  of bosonic and 
fermionic zero-mode norms and exactly matches the power expected on the
 basis of the AdS/CFT correspondence.   

The computation of the one-instanton contribution to the 
SYM correlation function 
$G_{16} = \langle {\Lambda}(x_{1})\ldots {\Lambda}(x_{16})
\rangle$, where $\Lambda^{A} = Tr(F_{\mu\nu} \sigma^{\mu\nu} \lambda^{A})$ 
is the fermionic composite operator dual to the type IIB dilatino,
and its comparison with the D-instanton contribution  
to the dual type IIB amplitude has given the 
first truly dynamical test of the correspondence \cite{BGKR}.
Correlation functions of this kind are almost completely determined
by the systematics of fermionic zero-modes  
in the YM instanton background.   
Performing (broken) superconformal transformations 
on the instanton field-strength
\be
F_{\mu\nu}(\rho_0, x_0; x) =  K_{2}(\rho_0, x_0; x) \sigma_{\mu\nu}
\ee
yields the relevant gaugino zero-modes
\be
\lambda^{A} = {1\over 2} F_{\mu\nu} \sigma^{\mu\nu} \zeta^{A}
\label{confzm}
\ee
where $\zeta^{A} = \eta^{A} + x\cdot \sigma \bar\xi^{A}$, with 
$\eta^{A}$ and $\bar\xi^{A}$ constant Weyl spinors of opposite 
chirality.
The exact matching with the corresponding type IIB amplitude is quite 
impressive and somewhat surprising. Indeed the SYM computation
initially performed for an $SU(2)$ gauge group at weak coupling 
\cite{BGKR}, \ie in a regime
which is clearly far from the large $N$ limit at strong coupling,
has since then been extended to the 
$k=1$ instanton sector for $SU(N)$ and to any
$k$ in the large $N$ limit \cite{DKHMV}.
The resulting 16-point functions have the same dependence 
on the insertion points. In the large $N$ limit the overall 
coefficients of the dominant terms are those predicted by the analysis 
of type IIB D-instanton effects \cite{GG}.
For $N \neq 2$ all but the 16 
superconformal zero-modes (\ref{confzm}) are lifted 
by Yukawa interactions. Additional bosonic coordinates 
parameterizing $S^{5}$ appear in the large $N$ 
limit as bilinears in the lifted fermionic 
zero-modes.

Other correlators that are related by supersymmetry to the $\Lambda^{16}$ 
function and can thus saturate the 16 exact zero-modes have been computed
\cite{BGKR,DKHMV,GK}. 
Correlators that cannot absorb the exact zero-modes receive vanishing 
contributions. This is the case for two- and 
three-point functions of CPO's as well as for extremal and 
next-to-extremal correlators to which we now turn our attention.

The correlator of CPO's 
\begin{equation}
     G_{extr} =
    \langle 
   {\cal Q}^{(\ell)}(x) {\cal Q}^{(\ell_{1})}(x_{1}) \ldots 
    {\cal Q}^{(\ell_{n})}(x_{n}) \rangle  
    \label{qextrcorr}
\end{equation}
is said to be ``extremal''
when $\ell=\ell_{1}+\ell_{2}+\ldots+\ell_{n}$.
It is easy to check that (\ref{qextrcorr}) contains only 
one $SU(4)$ tensor structure so that computing (\ref{qextrcorr})
is equivalent to computing
\begin{equation}
    \langle 
    Tr[(\phi)^{\ell}(x)] 
    Tr[(\phi^{\dagger})^{\ell_{1}}(x_{1})] \ldots 
    Tr[(\phi^{\dagger})^{\ell_{n}}(x_{n})] \rangle 
     \label{extrcorr}
\end{equation}
where $\phi$ is $\phi^{I} = \varphi^{I} + i \varphi^{I+3}$, with, say, 
$I=1$.
The tree-level contribution corresponds to a 
diagram with $\ell$ lines exiting from the point $x$, which form $n$ 
different ``rainbows'' connecting $x$ to the points $x_{i}$, the $i$th 
rainbow containing $\ell_{i}$ lines. The result is schematically of 
the form
\be
G(x,x_{1},\ldots,x_{n}) = c(g,N) \prod_{i} (x-x_{i})^{-2\ell_{i}} \; .
\ee

The dual supergravity computation is very subtle in 
that the relevant AdS integrals are 
divergent but at the same time extremal trilinear coupling are 
formally vanishing \cite{LMRS}. 
If one carefully analytically continue the 
computation away from extremality \cite{LT3,AF3}, 
one finds a non-vanishing result 
of the same form as at tree-level in SYM theory \cite{DFMMRE}. 
One is thus lead 
to conjecture extremal correlators should satisfy a non-renormalization 
theorem of the same kind as two- and three-point functions of CPO's.
This has been tested both at one-loop and non-perturbatively \cite{BKE}.

At one loop, there are two sources of potential corrections. 
The first corresponds to the insertion of a vector line connecting 
chiral lines of the same rainbow. Its vanishing is in some sense 
related to the vanishing of one-loop corrections to two-point 
functions of CPO's. The second corresponds to insertion of a vector 
line connecting chiral lines belonging to different rainbows. Its vanishing 
is in the same sense as above related to the vanishing of one-loop 
correction to three-point functions of CPO's. 
The same analysis can be repeated step by step in the case of 
extremal correlators involving multi-trace operators in short 
multiplets \cite{BKE}. 

As far as the instanton contributions are concerned,
it is easy to check that (\ref{extrcorr}) cannot absorb the 
relevant 16 zero-modes. The induced scalar zero-modes read
\be
\varphi^{i} = {1\over 2}
\tau^{i}_{AB} \zeta^{A} F_{\mu\nu} \sigma^{\mu\nu} \zeta^{B}
\ee
and the 4 exact zero-modes with flavour $I=1$
could only be possibly absorbed at $x$. Since however 
$\zeta (x)^{4} = 0$, the non perturbative corrections to (\ref{extrcorr}) 
vanishes for any instanton number and for any gauge group 
in the leading semiclassical approximation.

Other~correlators, involving only one $SU(4)$
singlet projection, enjoy similar non-renormalization properties. 
Two- and three-point functions of CPO's belong to 
this class. The identification of $U(1)_{B}$-violating nilpotent 
super-invariants beginning at 
five points \cite{EHSSWE} prevents one from generically 
extending the same argument to 
higher-point functions. However the absence of the relevant nilpotent 
super-invariants for next-to-extremal correlators, with $\ell = (\sum_{i} 
\ell_{i}) - 2$,  
allows one to add them to the above list \cite{EHSSWE}. 
The absence of one-loop and 
instanton corrections in this case \cite{EPV} 
can be verified along the same line as for the extremal ones
\cite{BKE}. Supergravity computations confirm the weak coupling result 
\cite{DEPV} and suggest that near-extremal correlators, 
with $\ell = (\sum_{i} \ell_{i}) - 4$, satisfy  a
partial non renormalization of some sort \cite{EPSS}. 
The {\it a priori} independent 
contributions to a given correlation function are functionally related 
to one another. Functional relations of this form 
easily emerge in instanton computations \cite{BKRS1}. Some additional 
effort allows one to derive them in perturbation theory 
\cite{EHSSW1,EHSSW2,BKRS1}.

The dynamics of the theory is elegantly encoded in the four-point 
functions. The simplest ones
have been computed both at weak coupling, up to order $g^4$ 
\cite{BKRS1,BKRS2,EHSSW1,EHSSW2} as well as
in the semiclassical instanton approximation 
\cite{BGKR}, and at strong coupling 
from the AdS perspective \cite{LT4,DFMMR,AF}. At
short distance they generically display logarithmic behaviours that 
are to be interpreted in terms of anomalous dimensions.
At first sight 
this might seem surprising in a theory, such as ${\cal
N}=4$ SYM, that is known to be finite.
Indeed, operators which belong to short 
multiplets have
protected scaling dimensions and cannot contribute to the logarithmic 
behaviours. Completeness of
the operator product expansion (OPE) 
requires the inclusion of ``unprotected'' operators 
in addition to the ``protected'' ones.
Single-trace operators in Konishi-like multiplets \cite{AFZ,ANS}  
contribute to the
logarithms at weak coupling but are expected to decouple at strong 
coupling. On the contrary unprotected multi-trace operators that are 
holographically dual to
multi-particle states appear both at weak and at strong coupling 
since their anomalous dimensions are at most of order $1/N^{2}$ 
\cite{MAGOO,DFMMR,AFP,BKRS1,BKRS2,BKRS3}. 

To clarify the point in a simpler setting, 
consider the two-point function of a primary
operator of scale dimension $\Delta = \Delta^{^{(0)}} + \gamma$.  
In  perturbation theory $\gamma=\gamma (g_{_{YM}})$ 
is expected to be small and to admit an expansion  in the
coupling constant $g_{_{YM}}$. Expanding in $\gamma$ yields  
\bea
&& \hspace*{-0.8cm}
\langle {\cal O}^\dagger_\Delta (x) {\cal O}_\Delta (y)\rangle
= { a_\Delta \over (x-y)^{2\Delta^{^{(0)}}}}\left( 1 - \right.
\\
&& \hspace*{-0.8cm}
\left. \gamma \log 
[\mu^2 (x-y)^2] + {\gamma^2 \over 2}  
( \log [\mu^2 (x-y)^2] )^2 + \ldots \right)
\nonumber 
\label{defdelta}
\eea
Although the exact expression~(\ref{twopoint}) given above
is conformally invariant, 
at each  order in $\gamma$~ (or in $g_{_{YM}}$) (\ref{defdelta}) contains 
logarithms that are an artifact of the pertrubative expansion. 

Similar considerations apply to arbritary 
correlation functions. 
Assuming the convergence of the OPE, a four-point 
function of primary operators can be schematically expanded as
\bea 
&&\langle {\cal Q}_{A}(x) {\cal Q}_{B}(y) {\cal Q}_{C}(z) {\cal Q}_{D}(w) 
\rangle  = \\
&&\sum_K {C_{AB}{}^K (x-y,\partial_{y})\over (x-y)^{\Delta_A + 
\Delta_B-\Delta_K}}
{C_{CD}{}^K (z-w,\partial_{w})\over (z-w)^{\Delta_C + 
\Delta_D-\Delta_K}}
\nonumber \\
&&\langle {\cal O}_{K}(y) {\cal O}_{K}(w) \rangle \; ,
\label{doubleope} \nonumber
\eea
where $K$ runs over a (possibly infinite) 
complete set of primary operators. Descendants 
are implicitly taken into account by the presence of derivatives in the 
Wilson coefficients, $C$'s. To simplify formulae we assume that 
${\cal Q}$'s are protected 
operators,~\ie they have vanishing 
anomalous dimensions. In general the operators ${\cal O}_{K}$ may 
have anomalous dimensions, 
$\gamma_K$, so that $\Delta_K=\Delta^{^{(0)}}_K + \gamma_K$. Similarly 
$C_{IJ}{}^K = C^{^{(0)}}_{IJ}{}^K + \eta_{IJ}{}^K$.
Indeed, although three-point functions of single-trace CPO's
are not renormalised beyond tree level~\cite{DFS}, 
{\it a priori} nothing can be said concerning corrections to three-point
functions also involving unprotected operators.

Neglecting descendants and keeping the lowest order terms
in $\gamma$ and $\eta$ 
\begin{eqnarray}
&&  \hspace*{-0.8cm}
\langle {\cal Q}_{A}(x) {\cal Q}_{B}(y) {\cal Q}_{C}(z) 
    {\cal Q}_{D}(w) \rangle_{_{(1)}} = 
    \\
   && \hspace*{-0.8cm}
\sum_K
    {\langle {\cal O}_{K}({y}) {\cal O}_{K}({w}) 
    \rangle_{_{(0)}} \over (x-y)^{\Delta_A + 
    \Delta_B-\Delta^{^{(0)}}_K} (z-w)^{\Delta_C + 
    \Delta_D-\Delta^{^{(0)}}_K}} \nonumber \\
    && \hspace*{-0.8cm} \times \left[ 
    \rule{0pt}{18pt}\eta_{AB}{}^K 
    C^{^{(0)}}_{CD}{}^K + C^{^{(0)}}_{AB}{}^K \eta_{CD}{}^{K} + 
    \right. \nonumber\\
    && \hspace*{-0.8cm}
\left.
    {\gamma_K \over 2} C^{^{(0)}}_{AB}{}^K 
    C^{^{(0)}}_{CD}{}^K \log {(x-y)^2 (z-w)^2 \over 
    (y-w)^4 }\right] \; .
    \label{opex}\nonumber
\end{eqnarray}
whence one can extract both corrections
to OPE coefficients and anomalous dimensions.

For definiteness, let us consider the four-point function of 
the lowest CPO's in the ${\cal N}$=4 current 
multiplet\footnote{AdS computations have also produced four-point functions 
of the scalar $SU(4)$ singlets in 
the dilaton-axion sector \cite{LT4,DFMMR} that are less amenable to explicit 
computations at weak coupling. To the best of my knowledge the only 
concrete proposal at weak coupling 
has been made in~\cite{BKC} for the instanton 
contribution.}, defined by  
\begin{equation}
{\cal Q}_{{\bf 20'}}^{ij} = \tr( \varphi^i \varphi^j - {\delta^{ij} 
\over 6} 
\varphi_k \varphi^k) \; . 
\end{equation} 

Due to the lack of a manifestly ${\cal N}=4$ off-shell superfield formalism,
perturbative computations have to be either performed in components or 
in one of the two available off-shell superfield formalisms. 
Although the number of diagrams is typically larger in the ${\cal N}=1$ 
superfield approach \cite{BKRS1,BKRS2,BKRS3} 
its simplicity makes it more accessible than the less 
familiar ${\cal N}=2$ harmonic superpace \cite{EHSSW1,EHSSW2}. 
It is remarkable that up to some overall normalization 
factors depending on the (not always standard) conventions adopted 
the two results are in perfect quantitative 
agreement with one another and in qualitative agreement with the AdS 
predictions at strong coupling. 

Instead of computing the most general four-point function of lowest 
CPO's we simply display the one-loop result for illustrative purposes.
One of the six $SU(4)$ singlet projections reads
\bea
&& G_{H}(x_{1}, x_{1}, x_{3}, x_{4}) = 
\\
&&\langle (\phi^{1})^{2}(x_{1}) (\phi^{\dagger}_{1})^{2}(x_{2})
(\phi^{2})^{2}(x_{3})(\phi^{\dagger}_{2})^{2}(x_{4}) \rangle =
\nonumber 
\\
&&- { 2 g_{_{YM}}^2 N (N^2-1) \pi^2 
\over
(2 \pi)^{12}  x_{12}^2 x_{34}^2 x_{13}^2 x_{24}^2} \  B(r,s) \; ,   
\label{gq4} 
\nonumber
\eea
where $B(r,s)$ is a box-type integral that 
can be expressed as a 
combination of logarithms and dilogarithms as follows
\ba
&& \hspace*{-0.6cm}
B(r,s)=  
{1 \over \sqrt{p}}
\left\{ \ln{r}\ln{s}  \right.  
\nonumber \\
&& - \ln^{2}\left({r+s-1 -\sqrt {p} \over 2}\right) 
\nonumber \\ 
&& \hspace*{-0.6cm}
-2 {\rm Li}_2 \left({2 \over 1+r-s+\sqrt {p}}\right )
\\
&& \left. 
-2 {\rm Li}_2 \left({2 \over 1-r+s+\sqrt {p}}\right )\right\}   
\nonumber
\ea
As indicated, $B(r,s)$ depends only on the two
independent conformally invariant cross ratios 
\be
	r = {x_{12}^{2}x_{34}^{2} \over x_{13}^{2}x_{24}^{2}}
	\;  , \quad
	s = {x_{14}^{2}x_{23}^{2} \over x_{13}^{2}x_{24}^{2}}
	\; .
\label{crossrat}
\ee
and 
\be 
p = 1 + r^{2} + s^{2} - 2r - 2s - 2rs \; .  
\label{pdef}
\ee
Thanks to crossing symmetry, the only 
other {\it a priori} independent four-point function of lowest CPO's is
\bea
&&G_{V}(x_{1}, x_{1}, x_{3}, x_{4}) = 
\\
&&\langle (\phi^{1})^{2}(x_{1}) (\phi^{\dagger}_{1})^{2}(x_{2})
(\phi^{1})^{2}(x_{3})(\phi^{\dagger}_{1})^{2}(x_{4}) \rangle
\nonumber
\eea
The non-perturbative contributions, computed in \cite{BGKR,BKRS1}, 
are quite involved and we refrain to display them. We simply 
notice that the relation \cite{EPSS}
\bea
&&(x_{1} - x_{3})^{2} (x_{2} - x_{4})^{2} G_{H}(x_{1}, x_{2}, x_{3}, x_{4}) 
= \nonumber \\
&&(x_{1} - x_{4})^{2} (x_{2} - x_{3})^{2} G_{V}(x_{1}, x_{2}, x_{3}, x_{4})
\label{gvgh}
\eea
can be easily derived from the systematics of the fermionic zero-modes 
\cite{BKRS1}. Some additional 
effort allows one to derive it in perturbation theory 
\cite{EHSSW1,EHSSW2,BKRS1,BKRS2}.
The AdS computation is even more involved and the final 
result is quite uninspiring \cite{AF}. 

In order to extract some physics one has to perform an OPE analysis. 
Restricting for brevity our attention 
to the sectors ${\bf 1}$, ${\bf 20'}$, ${\bf 84}$, and ${\bf 105}$
the results can be summarized as follows\footnote{Recall that 
in addition to these irreps, ${\bf 20'}\times {\bf 20'}$
contains ${\bf 15}+ {\bf 175}$ in the antisymmetric part.}.

In the ${\bf 105}$ one finds only subdominant logarithms, 
consistently with the 
expected absence of any corrections to the dimension 
of protected single- and double-trace operators of dimension $\Delta = 
4$ in the ${\bf 105}$ \cite{BKRS1,BKRS2,AFP,DFMMR,DEPV}. 

In the ${\bf 84}$ channel, the dominant contribution at one and two loops 
is purely logarithmic and    
consistent with the exchange of the operator ${\cal K}_{\bf 84}$
in the Konishi multiplet. The absence of dominant logarithmic terms in 
the instanton as well as AdS results suggests confirms the
absence of any corrections to the dimension and trilinear of 
a protected operator $\hat{\cal D}_{\bf 84}$ of dimension 4, 
defined by subtracting the 
Konishi scalar ${\cal K}_{\bf 84}$ 
from the projection on the 
${\bf 84}$ of the naive normal ordered product of two ${\cal Q}_{\bf 
20'}$ \cite{BKRS1,BKRS2,AFP,DFMMR,DEPV}. 

In the ${\bf 20'}$ sector, there is no dominant logarithm suggesting 
a vanishing anomalous dimension for the unprotected operator 
$:{\cal Q}_{\bf 20'}{\cal Q}_{\bf 20'}:_{\bf 20'}$ \cite{AFP}. This 
striking result seems to be a consequence of the partial 
non-renormalization of four-point functions of lowest CPO's \cite{EPSS,AEPS}
summarized by (\ref{gvgh}) that is valid not only at each order 
in perturbation theory (beyond tree level!) but also  
non-perturbatively and at strong coupling (AdS). 
In order to disentangle 
the various scalar operators of naive dimension 4 
exchanged in this channel it is necessary to 
compute other independent four-point functions involving the insertions 
of the lowest Konishi operator ${\cal K}_{\bf 1}$ \cite{BKRS3}. 

The analysis of the singlet channel is very complicated by the 
presence of a large number of operators. In perturbation theory one 
has logarithmically-dressed double pole associated to the exchange 
of ${\cal K}_{\bf 1}$ with\footnote{Notice that the coupling constant 
$g_{_{YM}}$ is related to the one in \cite{BKRS1,BKRS2} by 
$2 g_{_{YM}}=g$. I thank H. Osborn for pointing out to me
the discrepancy with \cite{ANS}.}
\be
\gamma^{(1)}_{{\cal K}} = 3 {g_{_{YM}}^{2} N \over 4\pi^{2}}
\qquad
\gamma^{(2)}_{{\cal K}} = - 3 {g_{_{YM}}^{4} N^2 \over 16 \pi^{2}}
\ee
Non-perturbative and strong coupling results only show a logarithmic 
singularity that is associated to the exchange of some double-trace 
unprotected operator ${\cal O}_{\bf 1}$ with $\gamma \approx 1/N^{2}$. 

The picture that emerges is very interesting. In addition to protected 
single an multi-trace operators satisfying shortening conditions 
as well as single- and multi-trace operators in long multiplets there 
seems to be a new class of operators that, though not satisfying any 
known shortening condition, have vanishing anomalous dimensions. 

Konishi-like operators decouple both from non-perturbative (instanton) 
correlators as well as from the strong coupling AdS results but they 
represent the only available window on genuine string dynamics 
\cite{BKRS3}. The OPE algebra at strong coupling requires the 
inclusion of multi-trace operators of three kinds. Those dual to 
multi-particle BPS states, those dual to non BPS-states with 
gravitational corrections to their binding energy and those 
dual to non BPS states without mass corrections. 
A deeper understanding of the last two classes of operators would 
help clarifying profound issues in the AdS/CFT correspondence such as 
the string exclusion principle that is expected to take over at 
finite $N$ \cite{MAGOO}. 

More importantly, by deforming ${\cal N} =4$ 
SYM it is possible to flow from the superconformal point to 
phenomenologically more interesting gauge theories with a 
dynamically generated mass gap \cite{GPPZ,FGPW,PW,BDFP}. Some insight 
on RG flows can be gained by means of the open-closed string duality. 
The running of the gauge coupling can be associated to dilaton-like 
tadpoles \cite{BMT} much in the same way as chiral anomalies are 
associated to RR tadpoles \cite{BMA}.

It is a pleasure to thank the organizers 
of the D.V. Volkov Memorial Conference ``Supersymmetry and Quantum Field
Theory'' in Kharkov, July 2000, for the kind invitation and the 
pleasant and stimulating atmosphere. I would also like to 
thank O. De 
Wolfe, D. Freedman, M. Green, S. Kovacs, J. F. Morales, K. Pilch, G. 
Rossi, K. Skenderis and Ya. Stanev for several enjoyable collaborations 
on the subject of this talk. 
This work was partly supported 
by the EEC contract HPRN-CT-2000-00122,  
a PPARC grant and the INTAS project 991590.

\end{document}